\documentclass{aastex}
\usepackage{url}
\usepackage{spr-astr-addons,amssymb}

\RequirePackage{color}

\newcommand{\emaila}{mgieles@eso.org}

\newcommand{\dr}{\mbox{${\rm d}$}}
\newcommand{\dm}{\mbox{${\rm d}M$}}
\newcommand{\mup}{\mbox{$M_{\rm up}$}}
\newcommand{\lup}{\mbox{$L_{\rm up}$}}
\newcommand{\mmin}{\mbox{$M_{\rm min}$}}

\newcommand{\nabove}{\mbox{$N(M>M_{\rm crit};t<10^8\,{\rm yr})$}}
\newcommand{\nabovethree}{\mbox{$N(M>10^3\,\msun)$}}
\newcommand{\msun}{\mbox{$M_\odot$}}
\newcommand{\mcrit}{\mbox{$M_{\rm crit}$}}
\newcommand{\mmax}{\mbox{$M_{\rm max}$}}
\newcommand{\mmaxth}{\mbox{$M_{{\rm max, 3rd}}$}}
\newcommand{\lmax}{\mbox{$L_{\rm max}$}}

\newcommand{\nbin}{\mbox{$N_{\rm bin}$}}

\newcommand{\dndt}{\mbox{${\rm d}N/{\rm d}t$}}
\newcommand{\dndtabove}{\mbox{${\rm d}N(M>10^3\,\msun)/{\rm d}t$}}
\newcommand{\dndm}{\mbox{${\rm d}N/{\rm d}M$}}
\newcommand{\dndl}{\mbox{${\rm d}N/{\rm d}L$}}

\begin{document}

\title{What determines the mass of the most massive star cluster in a galaxy: statistics, physics or disruption?}
\slugcomment{Springer-Verlag}
\shorttitle{The maximum cluster mass}
\shortauthors{Gieles et al.}

\author{M. Gieles\altaffilmark{1}}
\email{\emaila}

\altaffiltext{1}{European Southern Observatory, Santiago 19, Chile}

\begin{abstract}
In many different galactic environments the cluster initial mass function (CIMF) is well described by a power-law with index $-2$. This implies a linear relation between the mass of the most
massive cluster (\mmax) and the number of clusters. Assuming a constant cluster formation rate and no disruption of the most massive clusters it also means that \mmax\ increases linearly with age when determining \mmax\ in logarithmic age bins. We observe this increase  in five out of the seven galaxies in our sample, suggesting that \mmax\ is determined by the size of the sample. It also means that massive clusters are very stable against disruption, in disagreement with the  mass independent disruption (MID) model presented at this conference. For the clusters in M51 and the Antennae galaxies the size-of-sample prediction breaks down around $10^6\,\msun$, suggesting that this is a physical upper limit to the masses of star clusters in these galaxies. In this method there is a degeneracy between MID and a CIMF truncation. We show how the cluster luminosity function can serve as a tool to distinguish between the two.
\end{abstract}


\section{The most massive star cluster from statistical considerations}
In many studies of young star clusters the power-law is an ever returning function that is used to describe almost everything, where the only thing that varies is the index.  The cluster initial mass function (CIMF) is one of the famous  power-laws, with its index being $-2$ (e.g. \citealt{1997ApJ...480..235E, 1999ApJ...527L..81Z, 2003A&A...397..473B, 2003dhst.symp..153W, 2003MNRAS.343.1285D}). It is often assumed that the CIMF is ``universal". But is that really true? In this contribution we investigate what the consequences are of a universal power-law CIMF with index $-2$. This function has several appealing  properties that we use to make predictions, in particular for the most massive cluster in a galaxy.

Lets consider the ``universal" scenario, in which all cluster masses are drawn from the same  CIMF:

\begin{equation}
\phi(M)=\frac{\dr N}{\dr M}=A\,M^{-\alpha}, \hspace{1cm} \alpha=2.
\label{eq:mf}
\end{equation}

For $\alpha\leq2$ the total mass diverges at high masses, so $\phi(M)$ has to be truncated at some upper value, which can be interpreted as a limit above which no clusters can exist. We will refer to this limiting mass as \mup\ and to the most massive cluster observed, i.e. the most massive cluster actually formed, as \mmax. By choosing $\mup>>\mmax$ the CIMF behaves as if it has no truncation and the value of \mmax\ is the result of sampling statistics.

\mmax\ depends on the constant $A$ in Eq.~\ref{eq:mf} and can be found by solving

\begin{equation}
\int_{\mmax}^{\mup} \phi(M)\dr M=1.
\label{eq:intone}
\end{equation}

For $\alpha>1$ and $\mup>>\mmax$ this results in  $A=(\alpha-1)\,M^{\,\alpha-1}_{\rm max}$ and $A=\mmax$ for $\alpha=2$.  The relation between \mup\ and \mmax\ depends on the number of clusters formed ($N$) which is proportional to $A$. The larger $N$, the closer the statistically probable \mmax\ will be to \mup.  The relation between  \mmax\ and $N$ can be found from

\begin{eqnarray}
N&=&\int_{\mmin}^{\mup} A\,M^{-\alpha}\,\dm \label{eq:ntot0}\\
	        &\simeq& \frac{M^{\alpha-1}_{\rm max}}{M^{\alpha-1}_{\rm min}} , \alpha>1\label{eq:ntot1}\\
	        &\simeq& \frac{\mmax}{\mmin}, \alpha=2\label{eq:ntot2}.
\end{eqnarray}

For a constant \mmin\ we see from Eq.~\ref{eq:ntot1} that \mmax\ scales with $N$ as 

\begin{equation}
\mmax\propto N^{1/(\alpha-1)}.
\label{eq:mmaxntot}
\end{equation}

Assuming a constant cluster formation rate (CFR), \dndt=constant, the number of clusters in logarithmic age bins, \nbin, increases linearly with age, since 
$\nbin(t)\propto\dr N/\dr\ln(t)=t\,\dndt$
and so $\nbin(t)\propto t$.
The same holds for  $\dr N/\dr\log(t)$ apart from an additional constant $\ln(10)$.
Therefore \mmax\ in logarithmic age bins scales with age as \citep{2003AJ....126.1836H} 

\begin{equation}
\mmax\propto\nbin^{1/(\alpha-1)}Ê\propto t^{1/(\alpha-1)}.
\label{eq:mmax}
\end{equation}

\section{Variations in the cluster formation rate and mass independent disruption}
The derivation of Eq.~\ref{eq:mmax} was based on a constant CFR and no disruption. \citet{2003AJ....126.1836H} already noticed that a power-law relation for the CFR with age ($\dndt\propto t^\eta$, with $\eta$ negative for a CFR that was lower in the past) would change the relation for \mmax\ to

\begin{equation}
 \mmax\propto t^{(1+\eta)/(\alpha-1)}.
\label{eq:mmaxcfr}
 \end{equation}

In a similar way we can add the effect of the {\it mass independent dissolution} (MID) model that was proposed by \citet{2005ApJ...631L.133F}, \citet{2006ApJ...650L.111C}, \citet{2007AJ....133.1067W} and Chandar (in these proceedings). These authors argue that 90\% of the clusters are destroyed each age dex which, combined with a constant CFR, results in $\dndt\propto t^{-1}$ for mass limited cluster samples. For other MID percentages we can write $\dndt\propto t^{-\lambda}$ with $0<\lambda<1$ for MID percentages between 0\% and 90\%.
Adding MID to Eq.~\ref{eq:mmaxcfr} we get
\begin{equation}
 \mmax\propto t^{(1+\eta-\lambda)/(\alpha-1)}.
\label{eq:mmaxtot}
 \end{equation}

From Eq.~\ref{eq:mmaxtot}  it is visible that there is a degeneracy between MID and a CFR that has been increasing from the past to now ($\eta<0$). Because of the mass independent nature of this disruption model it is impossible to distinguish between an increasing CFR and MID.
For the universal 90\% MID scenario, which was proposed by \citet{2007AJ....133.1067W}, the predicted slope in a log(\mmax) {\it vs.} log(age) diagram is 0 for all values of $\alpha$ in all galaxies.  

If $\mmax=\mup$, i.e. a physical maximum to the cluster masses that is reached at all ages, \mmax\ is also constant with age. 
Hence there is also a degeneracy between the 90\% MID model and a truncation of the CIMF. In \S~\ref{sec:lf} we will show that the luminosity function is different in these two situations. In the following section we will compare empirical cluster ages and masses to the predictions for $\mmax(t)$.

\section{Comparison to observations}
\label{sec:obs}

\begin{figure*}[t]
   \centering
   \includegraphics[width=15.8cm]{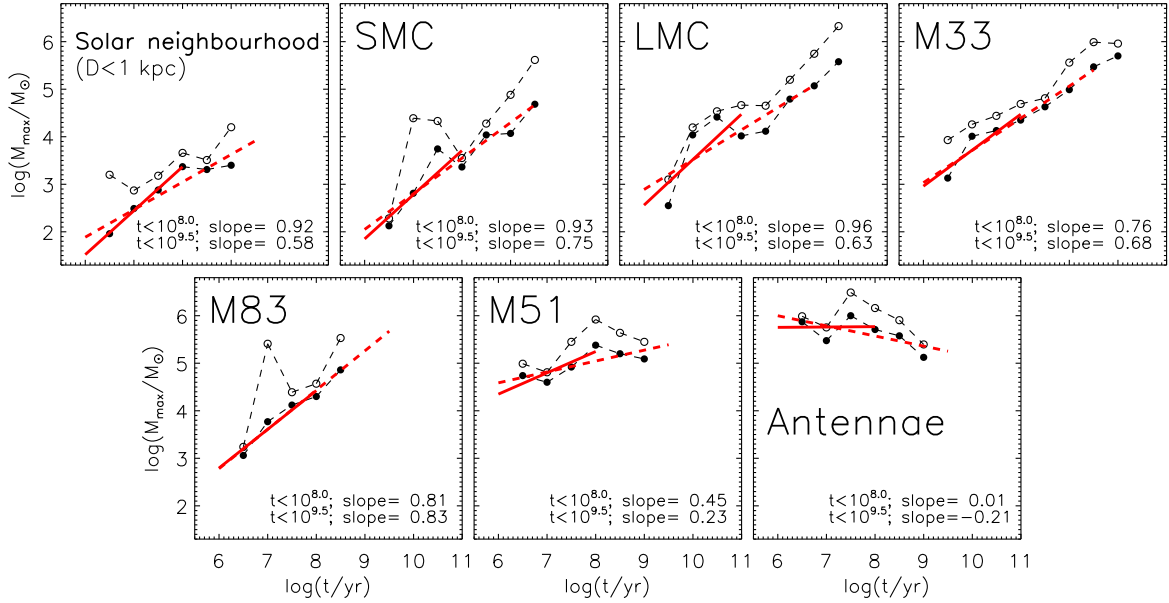}
   \caption{Evolution of log(\mmax) in log(age) bins for the seven galaxies in our sample (open circles). The third most massive clusters (\mmaxth) in each bin are shown as filled circles. Fits to $\log(\mmaxth)$ {\it vs.} log(age) on two different age ranges are shown as dashed and full lines.}
                    \label{fig:max}%
    \end{figure*}

\subsection{Description of the data used}
\label{subsec:obs}
From literature we collect ages and masses of star clusters in seven different galaxies:  the Milky Way (solar neighbourhood), SMC, LMC, M33, M83, M51 and the Antennae galaxies. The data were taken from  \citet{2005A&A...441..117L, 2005A&A...438.1163K}, \citet{2003AJ....126.1836H}, Mora et al. (2007, in prep), \citet{2005A&A...431..905B} and \citet{1999ApJ...527L..81Z}, respectively. We refer to \citet{gb07} for more details on the data.

\subsection{Observed trends of $\log(\mmax)$ {\it vs.} $\log$(age)}

In Fig.~\ref{fig:max} we show the trends of $\log(\mmax)$ with log(age) in bins of 0.5 dex as open circles. We also show the third most massive cluster in each age bin (\mmaxth) as filled circles,  which is expected to follow the same relation as \mmax, but with less scatter. 
The full lines are fits of $\log(\mmaxth)$ with log(age) over the first $100\,$Myr and the dashed lines consider an age range of 3 Gyr. 

For the solar neighbourhood, SMC, LMC, M33 and M83 the observed increase of $\log(\mmaxth)$ in the first $100\,$Myr is consistent with the size-of-sample prediction without disruption (Eq.~\ref{eq:mmax}). 
{\it This rules out the long term 90\% MID proposed by \citet{2007AJ....133.1067W}  as a universal cluster disruption mechanism for these galaxies. If the \citet{2007AJ....133.1067W} model was correct, \mmax\ had to be constant with log(age) for all galaxies. Fig.~\ref{fig:max} shows that this is clearly not the case.}

The trends for the interacting galaxies M51 and the Antennae are clearly different from the other five galaxies.  The flat slope beyond $100\,$Myr in M51 was interpreted as  a truncation of the CIMF  around $\mup\simeq5\times10^5\,\msun$ \citep{2006A&A...450..129G, 2006A&A...446L...9G}. We note that M83 was only partially imaged so  things could be different for this galaxy when all clusters are included since its CFR is probably similar to that of M51.

The flat relation for the Antennae galaxies is consistent with $90\%$ MID each age dex  during a Gyr. Note, however,  that if we attribute
 this result entirely to MID, i.e. we assume that the CFR has been roughly constant and that there is no truncation of the CIMF (the interpretation of \citealt{2007AJ....133.1067W}), then based on size-of-sample effects the Antennae galaxies should have formed clusters with masses up to $\sim10^9\,\msun$, but they have been destroyed by MID. This seems unlikely given what we know about other galactic mergers (i.e. that the major burst of star formation in the Antennae will happen as the nuclei coalesce, e.g. \citealt{2006MNRAS.373.1013C}).
 
\subsection{A universal cluster initial mass function?}

\begin{figure}[t]
   \centering
   \includegraphics[width=8cm]{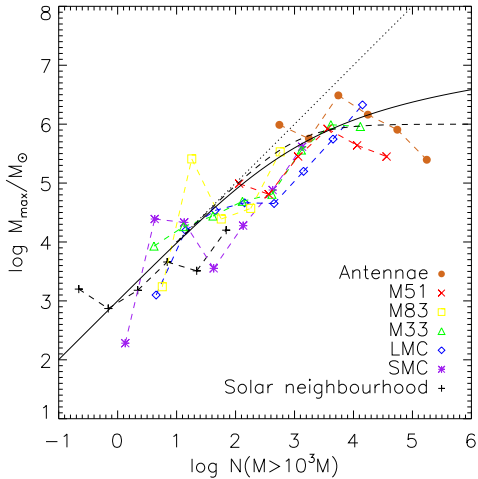}
   \caption{\mmax\ {vs.} \nabovethree\ for all galaxies from Fig.~\ref{fig:max}. The \nabovethree\ values for each bin of Fig.~\ref{fig:max} were 
   obtained by multiplying the bin widths ($\Delta t$) from Fig.~\ref{fig:max} by $\dndtabove$ (see the text for details). The \mmax\ values are the same as in Fig.~\ref{fig:max} for all galaxies. The dotted line represents a prediction for \mmax\ following from a power-law CIMF without a truncation (Eq.~\ref{eq:ntot2}). The dashed and full lines show a prediction based on a CIMF with $\mup=10^6\,\msun$ and a Schechter type CIMF with $M_*=10^6\,\msun$, respectively.}
   \label{fig:tot}
    \end{figure}

The galaxies in Fig.~\ref{fig:max} are (roughly) ordered by increasing star/cluster formation rate. The slope of \mmax\ with age is getting smaller going from the solar neighbourhood to the Antennae galaxies.  If cluster masses are drawn from a universal CIMF with a truncation at $\sim10^6\,\msun$ it is expected that for galaxies with a low CFR the CIMF is not sampled up to this mass ($\mmax<<\mup$)  and thus $\mmax$ follows from the size-of-sample prediction of Eq.~\ref{eq:mmax}. The truncation should then become more noticeable at high ages (larger age ranges) or in galaxies with a high CFR such as M51 and the Antennae galaxies.

We try to put the results of Fig.~\ref{fig:max}  in a unified \mmax\  {\it vs.} $N$ diagram, similar to what was done for the brightest cluster luminosity {\it vs.} $N$ \citep{2002AJ....124.1393L, 2003dhst.symp..153W, 2006A&A...450..129G}. Doing this for masses rather than luminosities is more difficult since we need to define $N$ as the number of clusters above a certain mass (\mcrit), with \mcrit\ the same for all galaxies and all age bins. However, \mmax\ in the nearby galaxies is of the same order as the minimum cluster mass we have available for M51 and the Antennae and the minimum observable cluster mass increases with age in samples that are luminosity limited (e.g. \citealt{2007ApJ...668..268G}).

From the available ages and masses we count the number of clusters with ages younger than $10^8\,$yr and masses above \mcrit, with $\log(\mcrit/\msun)=2$ for the solar neighbourhood, $\log(\mcrit/\msun)=3$ for the SMC and LMC, $\log(\mcrit/\msun)=3.5$ for M33 and M83, $\log(\mcrit/\msun)=4$ for M51 and $\log(\mcrit/\msun)=5$ for the Antennae. We then scale this number, \nabove, to a formation rate of clusters with masses above $10^3\,\msun$, \dndtabove, by dividing \nabove\ by $10^8\,$yr and multiplying it by $\mcrit/10^3$. The scaling with \mcrit\ is justified by Eq.~\ref{eq:ntot2}.  

We can now multiply the width of each bin of each galaxy in Fig.~\ref{fig:max} by the \dndtabove\ of that galaxy to get an estimate of the number of clusters with masses above $10^3\,\msun$ that has formed in each bin, again assuming a constant CFR. This allows us to plot all galaxies from Fig.~\ref{fig:max} in the same figure.
  
The result is shown in Fig.~\ref{fig:tot}. For  \mmax\ smaller than ${\sim10^6}\,\msun$  (and $\nabovethree\,\lesssim10^3$) all galaxies show the same linear increase of \mmax\ with \nabovethree, similar to that found for the brightest cluster luminosity. We plot the prediction for \mmax\ as a function of $N$ from Eq.~\ref{eq:ntot2}, i.e. for random sampling from a  $-2$ power-law CIMF, as a dotted line. There is good agreement with the observations for $\mmax\lesssim10^6\,\msun$. For higher \mmax\ the observed values flatten around $10^6\,\msun$. 

Using Eqs.~\ref{eq:intone}\&\ref{eq:ntot0} we solve for \mmax\ as a function of \nabovethree\ for $\mup=10^6\,\msun$ (dashed line) and we do the same for a Schechter type CIMF with $M_*=10^6\,\msun$ (full line). Both predictions including a truncation  describe the data very well suggesting that the masses of star clusters follow the size-of-sample prediction only up to $\sim10^6\,\msun$. Above that mass there are far fewer clusters observed than predicted  by random sampling from a power-law CIMF without a truncation.

\section{The cluster luminosity function as an independent check}
\label{sec:lf}

\begin{figure*}[t]
   \centering
   \includegraphics[width=12.cm]{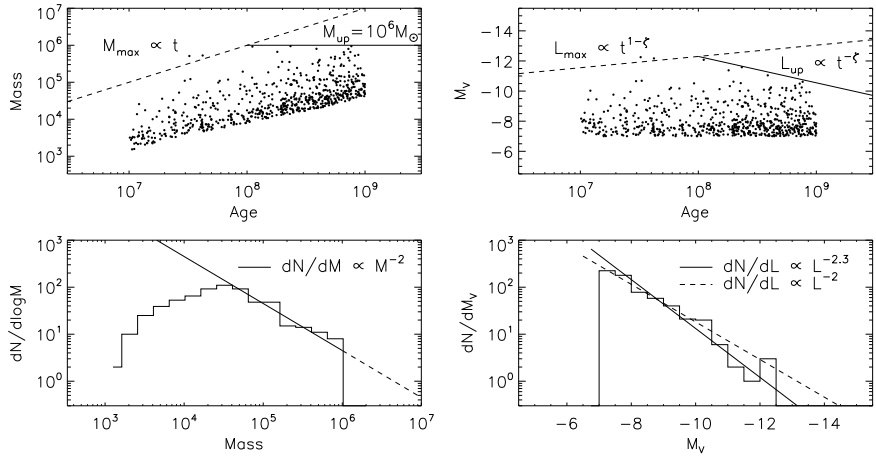}
   \caption{Simulated cluster population with $\dndtabove=10\,{\rm Myr}^{-1}$ and $\mup=10^6\,\msun$. {\it Top left:} Age-mass diagram of all cluster brighter than $M_V=-7$. {\it Top right:} Age-$M_V$ diagram. {\it Bottom left:} Mass distribution. {\it Bottom right:} Magnitude distribution. }
    \label{fig:panel}%
    \end{figure*}

Because the interpretation of the trends of log(\mmax) {\it vs.} log(age) could be affected by variations in the CFR and disruption (Eq.~\ref{eq:mmaxtot}), we show how the integrated luminosity function (LF) will look like when the CIMF is truncated.
When the mass function (MF) of a stellar population has a characteristic mass scale, then the LF of a multi-age population can not be interpreted as the MF due to the fading of clusters with age \citep{1995AJ....110.2665M,1997ApJ...488L..95H}. However, if all clusters are drawn from a power-law distribution with index $-2$, without a truncation, the integrated LF should still be a power-law with index $-2$. 
Age dependent extinction, star-bursts or variations in the CFR and mass {\it independent} dissolution will only affect the normalisation constant, but not the power-law index.  No matter how you normalise, a power-law plus a power-law is always a power-law. Mass {\it dependent} disruption will make the LF shallower, i.e. with an index larger than $-2$.

However, the LFs of multi-age cluster populations seem to have power-law indices that are {\it smaller} (between $-2.5$ and $-2$) (e.g. \citealt{2002AJ....123..207D, 2002AJ....123.1381E, 2002AJ....124.1393L}) with the LF being steeper at higher luminosities \citep{1999AJ....118..752Z, 2002AJ....123.1411B, 2002AJ....124.1393L, 1999AJ....118.1551W, 2005A&A...443...41M, 2006A&A...450..129G, 2006A&A...446L...9G, 2007arXiv0712.1420H}.

A slightly steeper LF results naturally from a truncated CIMF combined with evolutionary fading. If the CIMF is populated up to \mup\ at all ages, i.e. $\mmax=\mup$, then we have a constant number of the most massive clusters at all ages: $\dr N(\mup)/\dr t =$constant. The luminosity of these clusters (\lup), however, depends strongly on age. The light-to-mass ratio, or the flux of a cluster of constant mass, can be approximated as $t^{-\zeta}$, with $\zeta\simeq1.0/0.7/0.6$ for the $U/V/K$-band. Note that if $\mup>>\mmax$ then the luminosity of clusters with \mmax\ (\lmax) will actually (statistically) increase with age (in log(age) bins) as $\lmax\propto t^{1-\zeta}$. The fact that the most luminous cluster in a galaxy is generally young therefore already argues for a truncation of the CIMF or  a steepening at high masses. With $\lup\propto t^{-\zeta}$ we know the change of \lup\ with age 
$\dr\lup/\dr t\propto t^{-\zeta-1}$.
Combining this with the constant  $\dr N(\mup)/\dr t$ we can thus write
\begin{eqnarray}
\frac{\dr N}{\dr \lup}& \propto &t^{\zeta+1} \\
	                                 & \propto & L^{-1-1/\zeta}_{\rm up},\label{eq:dndlindex}
\end{eqnarray}
where $t$ is substituted for $L^{-1/\zeta}_{\rm up}$.

We thus predict that the bright end of the LF is a power-law with index $-1-1/0.7=-2.5$ in the $V$-band, whereas the faint end is a power-law with index $-2$. This is in perfect agreement with what was found by \citet{2006A&A...446L...9G} and \citet{1999AJ....118.1551W} for the LF of clusters in M51 and the Antennae\footnote{However, at this conference Brad Whitmore presented the LF of Antennae clusters derived from {\it HST/ACS} data and the break in the LF seems to have disappeared.}, respectively.
 In the $U$-band ($\zeta\simeq1$) the index of the LF is thus $-2$ and a truncation of the CIMF is not observable in the $U$-band LF. 
 This filter dependent power-law index is also found for the bright end of the LF of M51  clusters  \citep{2006A&A...450..129G, 2006A&A...446L...9G, haas}, though with low significance.

If \mmax\ is not sampled up to \mup\ at all ages, the index of the LF will be somewhere between $-2.5$ and $-2$. We illustrate this in Fig.~\ref{fig:panel} with a Monte Carlo simulation. We sample clusters from a CIMF with $\mmin=10^3\,\msun$ and $\mup=10^6\,\msun$, assuming a constant CFR between $10^7$yr and $10^9$yr and no disruption. We apply a detection limit at $M_V=-7$. In the top panels we show the masses and luminosities as a function of age.
For this CFR \mmax=\mup\ at $100\,$Myr, since $\mmax=\mmin*\dndtabove*\Delta t=10^3\times10\times100=10^6\,\msun$.  In the top left panel we illustrate the statistically expected \mmax\ as a function of log(age) as a dashed line. The points follow the line with some clusters above the line, which is allowed since \mmax\ is not  a hard limit. For ages older than $10^8\,$yr \mmax=\mup. 

In the bottom panels we show the resulting mass and luminosity distributions together with power-law approximations for $\dndm$ and $\dndl$. The mass function has a hard cut-off at high masses and is heavily affected by the detection limit at low masses. The LF, however, is still a power-law, but with a slightly smaller index ($\sim-2.3$). The LF is in fact best approached by a double power-law \citep{2006A&A...450..129G}, but in the situation where \mmax\ is not sampled up to \mup\ at all ages,  the LF becomes only slightly steeper than the underlying CIMF. Note that there are other effects that can make the LF even steeper,  but this is only when $\mmax\simeq\mup$ \citep{2002AJ....124.1393L}. We emphasise again that without a truncation the integrated LF should be a power-law with index $-2$.


\section{Conclusion}
The fact  that for  $\mmax\lesssim10^6\,\msun$ the relation between \mmax\ and \nabovethree\  is similar  in seven different galaxies argues for universality of the CIMF in that mass range. Whether the truncation mass  of $10^6\,\msun$ is universal is yet unclear. There are observational hints that \mup\ in the Antennae is a factor of four higher than in M51 \citep{2006A&A...450..129G, 2006A&A...446L...9G}. Moreover, star clusters with masses well in excess of $10^6\,\msun$ are known (e.g. \citealt{1997AJ....114.2381M, 2004A&A...416..467M, 2006A&A...448..881B, 2007A&A...469..147R, 2006ApJ...641..763W}). However, these objects seem to be different from ``regular" star clusters based on some fundamental properties. Above $\sim10^6\,\msun$ there is a relation between cluster radius and mass  \citep{2006A&A...448.1031K, 2005ApJ...627..203H}, while  below $\sim10^6\,\msun$ there in no such correlation, neither for young clusters \citep{2004A&A...416..537L, 2007A&A...469..925S}  nor for old clusters (e.g. \citealt{1994AJ....108.1292D, 2000ApJ...539..618M, 2005ApJ...634.1002J}). In addition, mass-to-light ratios increase above $\sim10^6\,\msun$ \citep{2007A&A...469..147R} and the chemical properties above this mass are distinct as well \citep{2006AJ....131.2442M}.
 Lastly, a universal upper mass helps to explain the universal globular cluster turn-over location \citep{2007ApJS..171..101J}.

\bibliographystyle{spr-mp-nameyear-cnd}


\end{document}